\documentclass[twocolumn,prb,showpacs,floatfix]{revtex4}

\usepackage{graphicx}
\usepackage{dcolumn}
\usepackage{bm}
\usepackage{amsmath}

\begin{document}

\title{
Structural properties of electrons in quantum dots in high magnetic fields:\\
Crystalline character of cusp states and excitation spectra
}

\author{Constantine Yannouleas}
\email{Constantine.Yannouleas@physics.gatech.edu}
\author{Uzi Landman}
\email{Uzi.Landman@physics.gatech.edu}

\affiliation{School of Physics, Georgia Institute of Technology,
             Atlanta, Georgia 30332-0430}

\date{17 August 2004; Revised 4 October 2004; 
Phys. Rev. B {\bf 70}, 235319 (2004)}

\begin{abstract}
The crystalline or liquid character of the downward cusp states in 
$N$-electron parabolic quantum dots (QD's) at high magnetic fields 
is investigated using conditional probability distributions obtained from 
exact diagonalization. These states are of crystalline character for 
fractional fillings covering both low and high values, unlike the liquid 
Jastrow-Laughlin wave functions, but in remarkable agreement with the 
rotating-Wigner-molecule ones [Phys. Rev. B {\bf 66}, 115315 (2002)]. 
The crystalline arrangement consists of concentric polygonal rings 
that rotate independently of each other, with the electrons on each ring
rotating coherently. We show that the rotation stabilizes the Wigner molecule 
relative to the {\it static\/} one defined by the broken-symmetry 
unrestricted-Hartree-Fock solution. We discuss the non-rigid behavior of the 
rotating Wigner molecule and pertinent features of the excitation spectrum,
including the occurrence of a gap between the ground and first-excited 
states that underlies the incompressibility of the system. This
leads us to conjecture that the rotating crystal (and not the static one) 
remains the relevant ground state for low fractional fillings even at the 
thermodynamic limit.
\end{abstract}


\maketitle

\section{Introduction}

The excitation energy spectrum of a two-dimensional $N$-electron semiconductor 
quantum dot (QD), plotted as a function of angular momentum
for a given high magnetic field $(B)$, exhibits downward cusps  
\cite{lau1,mak1,jai1,yl1,yl4b} at certain magic angular momenta ($L_m$), 
corresponding to states with enhanced stability. For a given value of $B$,
one of these $L_m$'s corresponds to the global minimum of the energy, that
is to the ground state (the ground-state value of $L_m$ is
denoted as $L_{\text{gs}}$). Varying the magnetic field causes the ground 
state and its angular momentum $L_{\text{gs}}$ to change. We note that due to
their enhanced stability, only cusp states can become ground states.
Underlying these properties is the inherent incompressibility of the cusp
states in response to an external magnetic field. As a result, the cusp states
have been long recognized \cite{lau1,jai1,haw,yl1,yl4b,yang} 
as the finite-$N$ precursors
of the fractional quantum Hall states in extended systems. In particular, the 
fractional fillings $\nu$ (defined in the thermodynamic limit) are related
to the magic angular momenta of the finite-$N$ system as follows: \cite{jach}
\begin{equation}
\nu = \frac{N(N-1)}{2L_m}.
\label{nu}
\end{equation}
(Henceforth, we will drop the subscript $m$, unless necessary).

In the literature of the fractional quantum Hall effect (FQHE), ever since the
celebrated paper \cite{lau2} by Laughlin in 1983, 
the cusp states have been considered to be the antithesis of the Wigner 
crystal and to be described accurately by liquid-like wave functions, such
as the Jastrow-Laughlin \cite{lau2,lau3} (JL) and composite-fermion 
\cite{jai2,jai4} (CF) ones. This view, however, has been recently challenged
\cite{yl1,yl4b} by the explicit derivation  of trial wave functions for the cusp 
states that are associated with a rotating Wigner (or electron) molecule, RWM. 
As we discussed \cite{yl1,yl4b} earlier, the parameter-free RWM wave functions, 
\cite{note11} which are by construction {\it crystalline\/} in character, 
promise to provide a simpler, but yet improved and more consistent description 
of the properties of the cusp states, in particular for high angular momenta 
(corresponding to low fractional fillings).

Issues pertaining to the liquid or crystalline character of the cusp states are
significant in both the fields of QD's and the FQHE. Since the many-body wave 
functions in the lowest Landau level (high $B$) obtained from exact 
diagonalization (EXD), the RWM wave functions, and the CF/JL ones have good 
angular momenta \cite{haw} $L \geq L_0=N(N-1)/2$, their electron densities are 
{\it circularly\/} symmetric. Therefore investigation of the crystalline or 
liquid character of these states requires examination of the conditional 
probability distributions (CPD's, i.e., the fully anisotropic pair correlation 
functions). These 
calculations were performed here under high magnetic field conditions for QD's 
(in a disk geometry \cite{note3}) with $N=6 - 9$ electrons, 
and for an extensive range of angular momenta. This allowed us to conclude 
that in all instances examined here (corrresponding to  $0.467>  
\nu > 0.111$) the cusp states exhibit an unmistakably crystalline character, 
in contrast to the long held perception in the FQHE literature, with the RWM 
yielding superior agreement with the exact-diagonalization results. \cite{yl6} 
Furthermore, the RWM states are found to be energetically stabilized 
(i.e., exhibit gain in correlation 
energy) with respect to the corresponding {\it static\/} (symmetry-broken) 
Wigner molecules, from which the multideterminantal RWM 
wavefunctions are obtained through an angular-momentum projection. \cite{yl4} 
We will present arguments that allow us to conjecture that the stabilization 
energy of the cusp states in high $B$ remains nonvanishing even in the 
thermodynamic limit.

In the beginning of sect.\ II, we display the Hamiltonian of the system
under consideration and define the conditional probability distributions.
Subsequently, in the same section, we present our main results pertaining to 
the structural properties of the CPD's. Possible improvementes of the RWM wave
functions are discussed in sect.\ II(c). In sect.\ III, we recapitulate the 
essential aspects of the two-step method of symmetry breaking and symmetry 
restoration, calculate stabilization energies for the RWM, and discuss 
pertinent features of its excitation spectrum (in particular, the occurrence 
of a gap between the ground state and the first excited state that is not a 
mere consequence of finite size; the appearance of this gap underlies the
incompressibility of the system). A discussion pertaining 
to implications for the thermodynamic limit is presented in sect.\ IV.
Finally, a summary is given in sect.\ V.

\section{Conditional probability distributions}

We are interested in wave functions which are exact solutions (or good 
approximations to them) of the two-dimensional many-body problem defined by the
Hamiltonian
\begin{equation}
H=\sum_{i=1}^N \frac{1}{2m^*}\left( {\bf p}_i- \frac{e}{c} {\bf A}_i \right)^2
+ \sum_{i=1}^N \frac{m^*}{2} \omega_0^2 {\bf r}_i^2
+ \sum_{i=1}^N \sum_{j > i}^N \frac{e^2}{\kappa r_{ij}},
\label{mbh}
\end{equation}
which describes $N$ electrons (interacting via a Coulomb repulsion) confined
in a parabolic potential of frequency $\omega_0$ and subjected to 
a perpendicular magnetic 
field $B$, whose vector potential is given in the symmetric gauge by
\begin{equation}
{\bf A(r)} = \frac{1}{2} {\bf B} \times {\bf r} = \frac{1}{2} (-By,Bx,0);
\label{vecp}
\end{equation}
$m^*$ is the effective electron mass, $\kappa$ is the  dielectric 
constant of the semiconductor material, and $r_{ij}=|{\bf r}_i - {\bf r}_j|$.
For sufficiently high magnetic field values (i.e., in the FQHE regime),
the electrons are fully spin-polarized and the Zeeman term (not shown here)
does not need to be considered.

In the $B \rightarrow \infty$ limit, the external confinement $\omega_0$ can 
be neglected, and $H$ can be restricted to operate in the lowest Landau level 
(LLL), reducing to the form:
\begin{equation}
H_{\text{LLL}} = N \frac{\hbar \omega_c}{2} 
+ \sum_{i=1}^N \sum_{j > i}^N \frac{e^2}{\kappa r_{ij}},
\label{hlll}
\end{equation}
where $\omega_c=eB/(m^*c)$ is the cyclotron frequency.

For finite $N$, the solutions to the Schr\"{o}dinger equations corresponding 
to the Hamiltonians (\ref{mbh}) and (\ref{hlll}) must have a good angular
momentum $L$. As described by us in detail in Refs. 
\onlinecite{yl1,yl4b,yl4,yl2} (see also sect. III 
below), these solutions can be well approximated by
a two-step method of symmetry breaking at the unrestricted Hartree-Fock (UHF)
level and of subsequent symmetry restoration via post-Hartree-Fock projection 
techniques. As elaborated in our earlier work, \cite{yl1,yl4b,yl4,yl2} 
the two-step method describes the formation and properties of rotating Wigner 
molecules in QD's.

As indicated in the Introduction, probing of structural characteristics in 
many-body wave functions with good angular momentum $L$ requires the use of the
conditional probability distributions defined by
\begin{equation}
P({\bf r},{\bf r}_0) =
\langle \Phi_L | 
\sum_{i=1}^N \sum_{j \neq i}^N  \delta({\bf r}_i -{\bf r})
\delta({\bf r}_j-{\bf r}_0) 
| \Phi_L \rangle  / \langle \Phi_L | \Phi_L \rangle.
\label{cpds}
\end{equation} 
Here $\Phi_L ({\bf r}_1, {\bf r}_2, ..., {\bf r}_N)$ 
denotes the many-body wave function under consideration.
In this paper, we calculate the CPD's for three types of many-body wave 
functions defined in the lowest Landau level:
(i) the analytic rotating Wigner molecule wave function, $\Phi^{\text{RWM}}_L$ 
(see also section III);
(ii) the wave function $\Phi^{\text{EXD}}_L$ obtained through exact 
diagonalization in the lowest Landau level; and (iii) the Jastrow/Laughlin 
functions $\Phi^{\text{JL}}_L$.

$P({\bf r},{\bf r}_0)$ is proportional to the conditional probability of
finding an electron at ${\bf r}$ under the condition that a second electron 
is located at ${\bf r}_0$. This quantity positions the observer on the moving
(intrinsic) frame of reference specified by the collective (coherent) 
rotations that are associated with the good angular momenta of the cusp states.

\begin{figure}[t]
\centering\includegraphics[width=6.5cm]{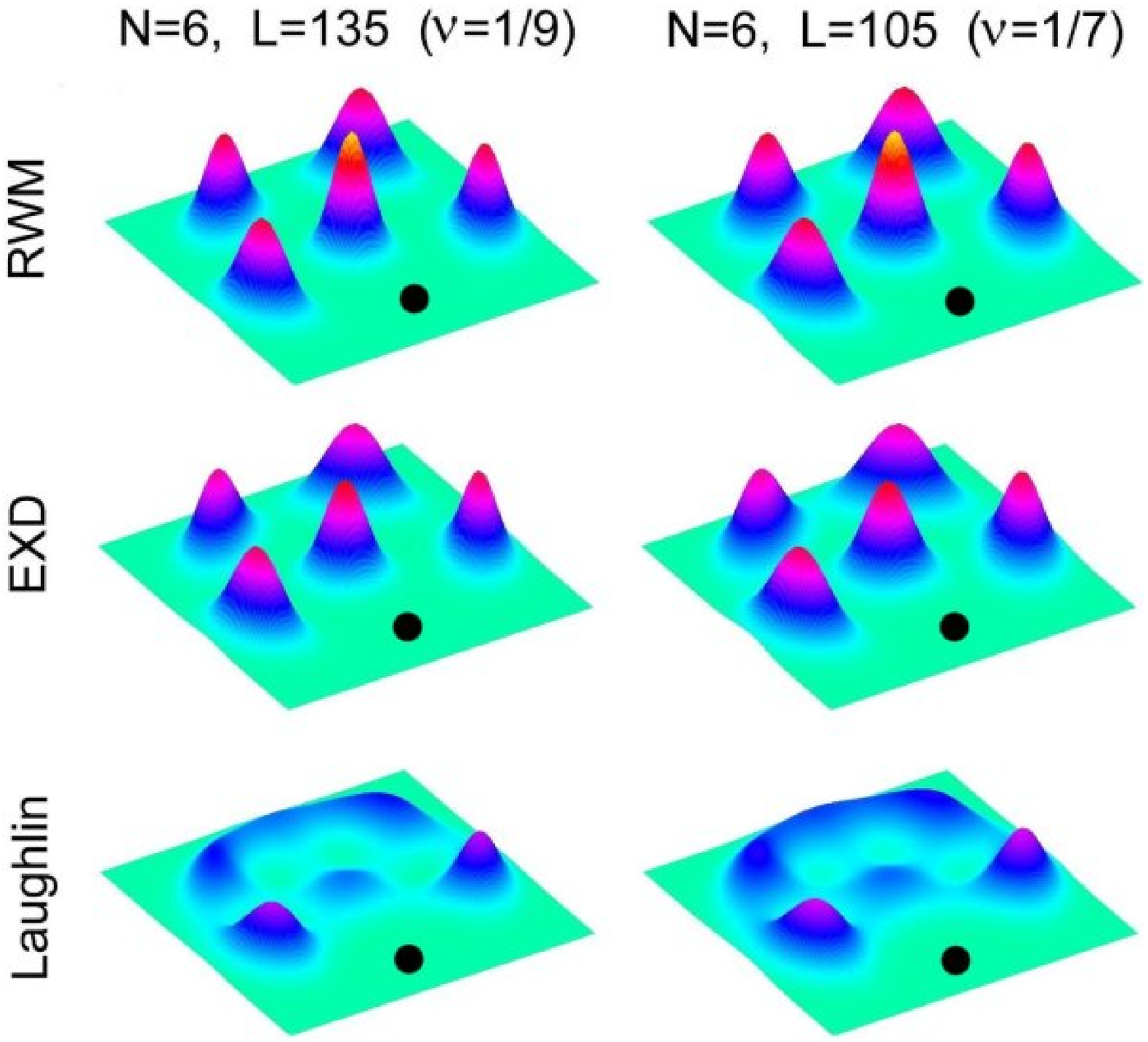}
\caption{
Conditional probability distributions at high $B$ for $N=6$ electrons with 
$L=135$ $(\nu=1/9$, left column) and $L=105$ $(\nu=1/7$, right column). 
Top row: RWM case. Middle row: The case of exact diagonalization. Bottom row:
The Jastrow-Laughlin case.
It is apparent that the exact diagonalization and RWM wave functions have a
pronouned crystalline character, corresponding to the (1,5) polygonal 
configuration of the rotating Wigner molecule. In contrast, the 
Jastrow-Laughlin wave functions fail to capture this crystalline character,
exhibiting a rather "liquid" character. The observation point 
(identified by a solid dot) was placed at the maximum of the 
outer ring of the radial electron density \cite{yl1,yl4b} of the EXD
wave function, namely at $r_0=7.318 l_B$ for $L=135$ and $r_0 = 6.442  l_B$ 
for $L=105$. Here, $l_B=(\hbar c/eB)^{1/2}$. 
The EXD Coulomb interaction energies (in the lowest Landau level)
are 1.6305 and 1.8533 $e^2/\kappa l_B$ for $L=135$ and $L=105$, respectively. 
The errors relative to the corresponding EXD 
energies and the overlaps of the trial functions with the EXD ones are: 
(I) For $L=135$, RWM: 0.34\%, 0.860; JL: 0.50\%, 0.665. 
(II) For $L=105$, RWM: 0.48\%, 0.850; JL: 0.46\%, 0.710. 
}
\end{figure}

\begin{figure}[t]
\centering\includegraphics[width=6.5cm]{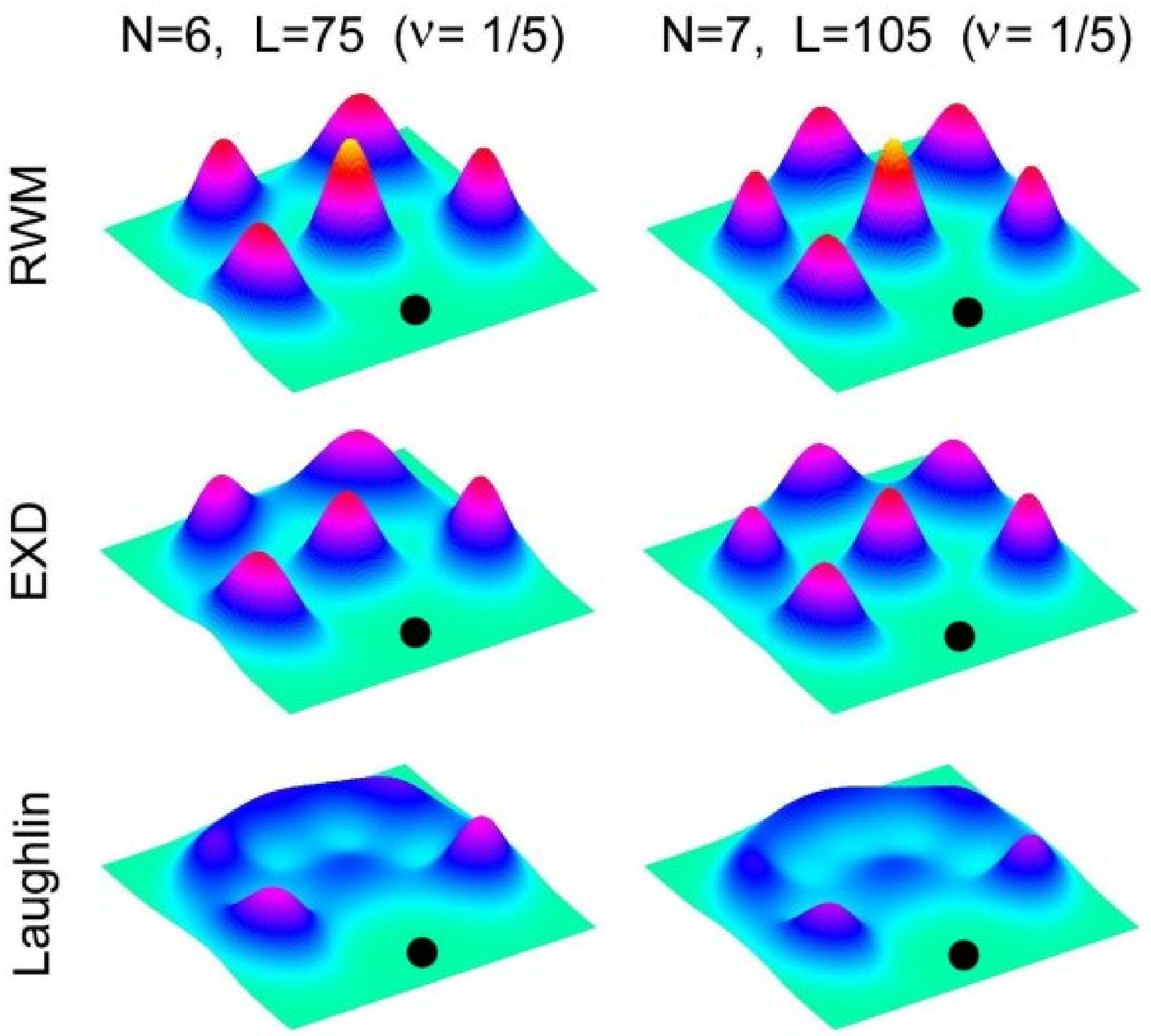}
\caption{
Conditional probability distributions at high $B$ for $N=6$ electrons and 
$L=75$ ($\nu=1/5$, left column) and for $N=7$ electrons and $L=105$ (again 
$\nu=1/5$, right column). Top row: RWM case. Middle row: The case of exact 
diagonalization. Bottom row: The Jastrow-Laughlin case.
The exact diagonalization and RWM wave functions have a pronouned crystalline
character, corresponding to the (1,5) polygonal configuration of the RWM for 
$N=6$, and to the (1,6) polygonal configuration for $N=7$. In contrast, the 
Jastrow-Laughlin wave functions exhibit a characteristic liquid profile that 
depends smoothly on the number $N$ of electrons. The observation point 
(identified by a solid dot) is located at $r_0 = 5.431 l_B$ 
for $N=6$ and $L=75$ and $r_0=5.883 l_B$ for $N=7$ and $L=105$.
The EXD Coulomb interaction energies (lowest Landau level)
are 2.2018 and 2.9144 $e^2/\kappa l_B$ for $N=6,L=75$
and $N=7,L=105$, respectively. The errors relative to the corresponding EXD
energies and the overlaps of the trial functions with the EXD ones are:
(I) For $N=6,L=75$, RWM: 0.85\%, 0.817; JL: 0.32\%, 0.837. 
(II) For $N=7,L=105$, RWM: 0.59\%, 0.842; JL: 0.55\%, 0.754. 
}
\end{figure}

\subsection{Crystallinity in lower fractions $ 1/9 \leq \nu \leq 1/5 $}

The CPD's for cusp states corresponding to a lower filling factor than
$\nu = 1/5$, calculated for $N=6$ electrons with $L=135$ ($\nu=1/9$, left 
column) and for $N=6$ with $L=105$ ($\nu=1/7$, right column) are displayed in
Fig.\ 1. Fig.\ 2 displays the CPD's for the cusp states with $N=6$ 
electrons and $L=75$ ($\nu=1/5$, left column) and $N=7$ and $L=105$ 
($\nu=1/5$, right column). In both figures, the top row depicts the RWM case. 
The EXD case is given by the middle row, while the CF cases (which reduce to 
the JL wave functions for these fractions) are given by the bottom row. 

There are three principal conclusions that can be drawn from an inspection
of Figs.\ 1 and 2: 

(I) The character of the EXD states is unmistakably crystalline with the EXD 
CPD's exhibiting a well developed molecular polygonal configuration [(1,5) for
$N=6$ and (1,6) for $N=7$, with one electron at the center], in agreement with 
the explicitly crystalline RWM case. 

(II) For all the examined instances (covering the 
fractional fillings 1/9, 1/7, and 1/5), the JL wave functions fail to capture 
the intrinsic crystallinity of the EXD states. In contrast, they represent
``liquid'' states in agreement with an analysis that goes back to 
the original papers \cite{lau2,lau3} by Laughlin. In particular, Ref.\ 
\onlinecite{lau3} investigated the character of the JL states through the use 
of a pair correlation function [usually denoted by $g(R)$] that determines the 
probability of finding another electron at the absolute relative distance 
$R=|{\bf r} - {\bf r}_0|$ from the observation point ${\bf r}_0$. 
Our anisotropic CPD of Eq.\ (\ref{cpds}) is of course more general (and more 
difficult to calculate) than the $g(R)$ function of Ref.\ \onlinecite{lau3}. 
However, both our $P({\bf r},{\bf r}_0)$ (for $N=6$ and $N=7$ electrons) and 
the $g(R)$ (for $N=1000$ electrons, and for $\nu=1/3$ and $\nu=1/5$) in Ref.\ 
\onlinecite{lau3} reveal a similar characteristic liquid-like and
short-range-order behavior for the JL states, eloquently described in Ref.\ 
\onlinecite{lau3} (see p. 249 and p. 251). Indeed, we remark that only the 
first-neighbor electrons on the outer rings can be distinguished as separate 
localized electrons in our CPD plots of the JL functions [see Fig.\ 1 and 
Fig.\ 2]. 

(III) For a finite number of electrons, pronounced crystallinity of the EXD 
states occurs already at the rather high $\nu=1/5$ value (see Fig.\ 2). This
finding is particularly interesting in light of expectations \cite{jai3}
(based on comparisons \cite{lau2,lau3,lam} between the JL states and the 
static bulk Wigner crystal) that a liquid-to-crystal phase transition may take
place only at lower fillings with $\nu \leq 1/7$. 

\begin{figure}[t]
\centering\includegraphics[width=6.5cm]{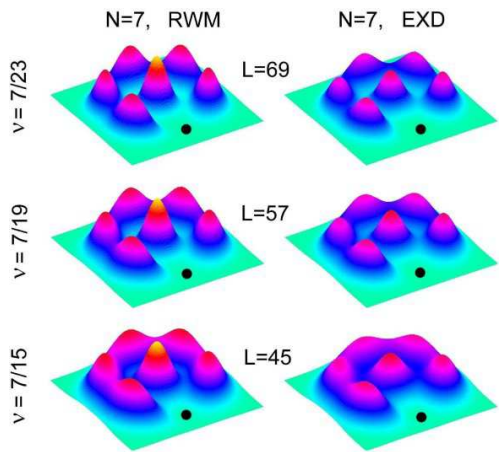}
\caption{
Conditional probability distributions at high $B$ for $N=7$ electrons and
$L=69$ ($\nu=7/23=0.304 > 1/5$, top row), $L=57$ ($1> \nu=7/19=0.368 > 1/3$,
middle row), and $L=45$ ($1> \nu=7/15=0.467 > 1/3$, bottom row). RWM case:
Left column. The case of exact diagonalization is depicted in the right
column. Even for these low magic angular momenta (high fractional fillings),
both the exact-diagonalization and RWM wave functions have a pronouned
crystalline character [corresponding to the (1,6) polygonal configuration of
the RWM for $N=7$ electrons]. The observation point (identified by a solid 
dot) is located at $r_0=4.752 l_B$ for $L=69$, $r_0=4.278 l_B$
for $L=57$, and $r_0=3.776 l_B$ for $L=45$.
}
\end{figure}

\subsection{Crystallinity in higher fractions $ 1/5 < \nu < 1 $}

Following the conclusion that the crystalline character of the cusp states in
QD's is already well developed for fractional fillings with the unexpected high
value of $\nu=1/5$, a natural question arises concerning the presence or
absence of crystallinity in cusp states corresponding to higher fractional
fillings, i.e., states with $ 1/5 < \nu < 1/3$, and even with $ 1/3 < \nu < 1$.
To answer this question, we show in Fig.\ 3 the CPD's for the RWM
(left column) and EXD (right column) wave functions for the case of $N=7$
electrons and for $L=69$ ($1/5 < \nu=7/23 < 1/3$), $L=57$ ($1/3 < \nu=7/19
< 1$), and $L=45$ ($1/3 < \nu=7/15 < 1$). Unlike the long held 
perceptions in the FQHE literature (which were reasserted in two recent 
publications \cite{jai3}) the CPD's in Fig.\ 3 demonstrate that 
the character of the cusp states with high fractional fillings is not 
necessarily ``liquid-like''. Indeed, these high-$\nu$ cusp states exhibit a 
well developed crystallinity associated with the (1,6) polygonal 
configuration of the RWM, appropriate for $N=7$ electrons.

\begin{figure}[t]
\centering\includegraphics[width=4.cm]{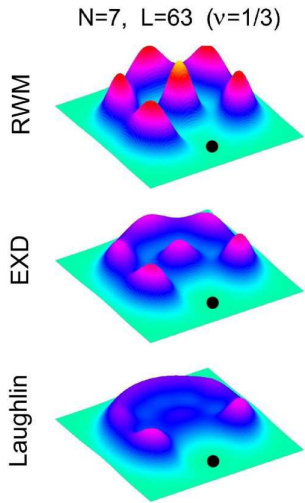}
\caption{
CPD's at high $B$ for $N=7$ and $L=63$ ($\nu=1/3$). 
Top: RWM case; Middle: EXD case; 
Bottom: JL case. Unlike the JL CPD (which is liquid), the CPD's for
the exact-diagonalization and RWM wave functions exhibit a well developed
crystalline character [corresponding to the (1,6) polygonal 
configuration of the RWM for $N=7$ electrons]. The observation point 
(identified by a solid dot) is located at $r_0=4.568 l_B$.
}
\end{figure}

Of interest also is the case of $\nu=1/3$. Indeed, for this
fractional filling, the liquid JL function is expected to provide the best
approximation, due to very high overlaps (better than 0.99) with the exact 
wave function. \cite{yosh2,tsip} In Fig.\ 4, we display the CPD's
for $N=7$ and $L=63$ ($\nu=1/3$), and for the three cases of RWM, EXD, and JL
wave functions. Again, even in this most favorable case, the CPD of the
JL function disagrees with the EXD one, which exhibits clearly a (1,6) 
crystalline configuration in agreement with the RWM CPD. 

Similar crystalline correlations at higher fractions were also found for QD's 
of a different size, e.g., with $N=6$, $N=8$, and $N=9$ electrons. As 
illustrative examples for these additional sizes, we 
display in Fig.\ 5 the CPD's for $N=8$ and $L=91$ ($1/5 < \nu=4/13 < 1/3$)
and for $N=9$ and $L=101$ ( $1/3 < \nu=36/101 < 1$). Again, the CPD's
(both for the RWM and the EXD wave functions)
exhibit a well developed crystalline character in accordance with the 
(1,7) and (2,7) polygonal configurations of the RWM, appropriate for $N=8$ and
$N=9$ electrons, respectively.

The case of $N=9$ is of particuler significance. Indeed it represents the 
smallest number of electrons with a non-trivial concentric-ring arrangement, 
i.e., the inner ring has more than one electrons. As the two CPD's (reflecting
the choice of taking the observation point [${\bf r}_0$ in Eq.\ ({\ref{cpds})]
on the outer or the inner ring) for $N=9$ reveal, the polygonal electron rings
rotate {\it independently\/} of each other. Thus, e.g., to an observer located
on the inner ring, the outer ring will appear as uniform, and vice versa.
The fact that both the RWM and exact wave functions share this property of
independently rotating rings is a testament to the ability of the RWM theory
to capture the essential physics of QD's in high $B$.

\begin{figure}[t]
\centering\includegraphics[width=6.5cm]{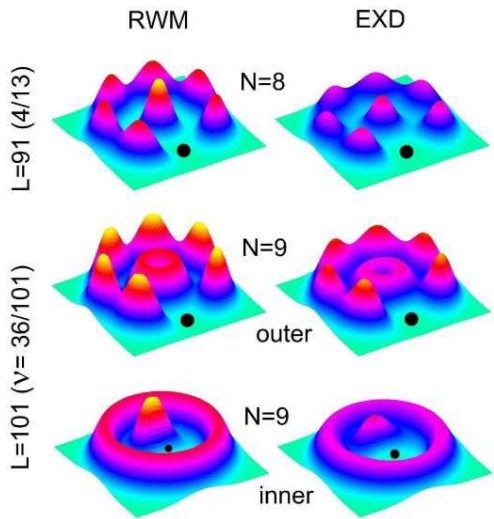}
\caption{
Additional CPD's at high $B$. RWM results: Left column. Results from exact 
diagonalization are depicted on the right column. 
Top row: $N=8$ electrons and $L=91$ 
($1/5 < \nu=4/13=0.308 < 1/3$).
Two bottom rows: $N=9$ electrons and $L=101$ 
($1/3 < \nu=36/101=0.356 < 1$, see text for explanation).
Even for these low magic angular momenta (high fractional fillings), 
both the exact-diagonalization and RWM wave functions have a pronouned 
crystalline character [corresponding to the (1,7) and (2,7) polygonal 
configurations of the RWM for $N=8$ and 9 electrons].
The observation point (identified by a solid dot) is located at 
$r_0=5.105 l_B$ for $N=8,L=91$, and $r_0=5.218 l_B$ (outer) and 
$r_0=1.662 l_B$ (inner) for $N=9,L=101$. 
}
\end{figure}

\subsection{Improvements of the RWM wave functions}

It is of importance to note here that the favorable comparison between the
crystalline structure of the RWM and that of the exact wave functions, in
contrast to the liquid-like character of the JL functions, persists even
for cases where the latter is found to have the advantage (over the RWM)
concerning total energies and wave function overlaps. As examples, we refer to
the case of $\nu =1/5$ discussed in the caption of Fig.\ 2 (see in particular
the errors in the energies for the RWM and JL functions relative to the exact
energies given at the end of the caption).

Close inspection of the humps in the CPD's obtained from the RWM and through
exact diagonalization reveals that the RWM tends to somewhat 
overestimate the degree of crystallinity, i.e., the RWM humps are narrower and 
higher (this tendency diminishes for larger values of $L$). Nevertheless,
the degree of overall agreement between the exact results and those obtained
through the {\it parameter-free\/} RWM wave functions is rather remarkable.
Moreover, the high level of agreement between the RWM and exact results 
extends to other properties. This includes the zeroes (often called vortices)
of the many-body wave functions. Indeed, as recently shown in Ref. 
\onlinecite{anis}, the exact wave functions (in contrast to the JL ones) have 
{\it simple zeroes\/} whose topology is in agreement with that of 
the simple zeroes of the RWM functions. 

The above suggest that the RWM wave functions can form the nucleus for
constructing a whole class of rotating crystalline functions with added
variational freedom, which will yield further quantitative energetic and 
structural improvements. For example the RWM functions could be used as the 
basis for constructing variational wave functions in diffusion \cite{bolt} and
variational \cite{nie} quantum Monte-Carlo studies. For a most recent
investigation along these lines, see Ref.\ \onlinecite{jai5}, 
where our RWM function is augmented by 
a Jastrow prefactor with an exponent that is treated variationally.
We remark, however, that the variational wave function employed in Ref.\ 
\onlinecite{jai5} has multiple zeroes due to the Jastrow factor, in 
disagreement with the exact diagonalization results.

\section{Restoration of circular symmetry}

\subsection{Correlated many-body wave functions}

Our two-step method for deriving the RWM wave functions is anchored in 
the distinction \cite{yl2} between a {\it static\/} and a {\it rotating} Wigner
molecule, with the rotation stabilizing the latter relative to the 
former. Further elaboration on this point requires generation of global ground 
states out of the cusp states, achieved through inclusion \cite{yang,yl2} of an
external parabolic confinement (of frequency $\omega_0$).
In the two-step method, the static WM is first described by an unrestricted 
Hartree-Fock (UHF) determinant that violates the circular symmetry. \cite{yl3} 
Subsequently, the rotation of the WM is described by a  
post-Hartree-Fock step of restoration of the broken circular symmetry via 
projection techniques. \cite{yl4} We note that, in the limit 
$N \rightarrow \infty$, the static WM of the UHF develops to the extended
two-dimensional Wigner crystal \cite{yosh} and its more sophisticated 
variants. \cite{lam} 

In general, the localized broken symmetry orbitals of the HF determinant are 
determined numerically via a selfconsistent solution of the UHF equations. 
Since we focus here on the case of high $B$, we can approximate 
the UHF orbitals (first step of our procedure) by (parameter free) displaced 
Gaussian functions; namely, for an electron localized at ${\bf R}_j$
($Z_j$), we use the orbital
\begin{equation}
u(z,Z_j) = \frac{1}{\sqrt{\pi} \lambda}
\exp \left( -\frac{|z-Z_j|^2}{2\lambda^2} - i\varphi(z,Z_j;B) \right),
\label{uhfo}
\end{equation}
with $\lambda = \sqrt{\hbar /m^* \Omega}$;
$\Omega=\sqrt{\omega_0^2+\omega_c^2/4}$, where $\omega_c=eB/(m^*c)$ is the
cyclotron frequency. We have used complex numbers to represent the position
variables, so that $z=x+iy$, $Z_j = X_j +i Y_j$.
The phase guarantees gauge invariance in the presence of
a perpendicular magnetic field and is given in the symmetric gauge by
$\varphi(z,Z_j;B) = (x Y_j - y X_j)/2 l_B^2$, with $l_B = \sqrt{\hbar c/ e B}$.
We only consider the case of fully polarized electrons, which is appropriate at
high $B$. 

We take the $Z_j$'s to coincide with the equilibrium positions [forming a
structure of $r$ concentric regular polygons 
denoted as ($n_1, n_2,...,n_r$)] of 
$N=\sum_{q=1}^r n_q$ classical point charges inside an external parabolic 
confinement of frequency $\omega_0$. Then we proceed to construct the UHF 
determinant $\Psi^{\text{UHF}} [z]$ out of the orbitals $u(z_i,Z_i)$'s, 
$i = 1,...,N$. Correlated many-body states with good total angular momenta $L$
can be extracted \cite{yl4b,yl4} from the UHF determinant using projection 
operators, i.e.,
\begin{eqnarray}
\Phi^{\text{RWM}}_L & = & \int_0^{2\pi} ... \int_0^{2\pi} 
d\gamma_1 ... d\gamma_r \nonumber \\
&& \times \Psi^{\text{UHF}}(\gamma_1, ..., \gamma_r) 
\exp \left( i \sum_{q=1}^r \gamma_q L_q \right),
\label{wfprj}
\end{eqnarray}
where $L=\sum_{q=1}^r L_q$ and
$\Psi^{\text{UHF}}[\gamma]$ is the original UHF determinant with {\it all
the orbitals of the $q$th ring\/} rotated (collectively, i.e., coherently) by 
the {\it same\/} azimuthal angle $\gamma_q$, and each ring is rotated
independently of each other. Note that Eq.\ (\ref{wfprj}) can be
written as a product of projection operators acting on the original UHF
determinant [i.e., $\Psi^{\text{UHF}}(\gamma_1=0, ..., \gamma_r=0)$,
see Eqs.\ (6) and (7) in Ref.\ \onlinecite{yl4b}]. 
Setting $\lambda = l_B \sqrt{2}$ restricts the orbital in Eq. 
(\ref{uhfo}) to lie entirely in the lowest Landau level, and allows for
the derivation of the analytic RWM functions. \cite{yl4b}
We stress that while the initial trial wave function of the UHF equations
consists of a single determinant, the projected wave function is a linear
superposition of many determinants, as can be explicitly seen from the
analytic forms of the RWM functions in Refs. \onlinecite{yl4b}.

\subsection{Stabilization energy}

In the case of finite $B$ (requiring the inclusion of confinement, i.e.,
$\omega_0 \neq 0$), the projected energy corresponding to a
symmetry-restored RWM state with angular momentum $L$ is given [in the case
of a single $(0,N)$ or a $(1,N-1)$ ring] by \cite{yl4,yl2}
\begin{equation}
E_{\text{PRJ}} (L) = \left. { \int_0^{2\pi} h(\gamma) e^{i \gamma L}
d\gamma } \right/ { \int_0^{2\pi} n(\gamma) e^{i \gamma L} d\gamma},
\label{eproj}
\end{equation}
with $h(\gamma) =
\langle \Psi^{\text{UHF}}(0) | H | \Psi^{\text{UHF}}(\gamma) \rangle$
and
$n(\gamma) =
\langle \Psi^{\text{UHF}}(0) | \Psi^{\text{UHF}}(\gamma) \rangle$,
where $\Psi^{\text{UHF}}(\gamma)$ is the original UHF determinant with {\it all
the orbitals\/} rotated (collectively) by the {\it same\/} 
azimuthal angle $\gamma$. $H$ is the many-body Hamiltonian in Eq. (\ref{mbh}). 
The UHF energies are simply given by $E_{\text{UHF}} = h(0)/n(0)$.

\begin{figure}[t]
\centering\includegraphics[width=7.0 cm]{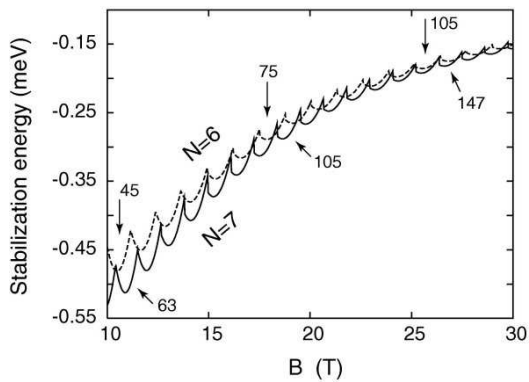}
\caption{Stabilization energies $\Delta E^{\text{gain}}_{\text{gs}}$
for $N=6$ (dashed curve) and 
$N=7$ (solid curve) fully polarized electrons in a parabolic QD as a function 
of $B$. The troughs associated with the major fractional fillings (1/3, 1/5, 
and 1/7) and the corresponding ground-state angular momenta 
[see Eq.\ (\ref{nu})] are indicated with
arrows. We have extended the calculations up to $B=120$ T (not shown), and 
verified that $\Delta E^{\text{gain}}_{\text{gs}}$ remains negative while its 
absolute value vanishes as $B \rightarrow \infty$.
The choice of parameters is: $\hbar \omega_0=3$ meV (parabolic confinement), 
$m^*=0.067 m_e$ (electron effective mass), and $\kappa =12.9$ (dielectric
constant).
}
\end{figure}

We note that, unlike the UHF ground state (describing a static Wigner 
molecule) which does not have good angular momentum, the ground states of the
RWM exhibit good angular momenta (labeled as $L_{\text{gs}}$, 
as aforementioned) that coincide with magic ones [we denote the 
ground-state energy of the RWM as 
$E^{\text{gs}}_{\text{PRJ}} \equiv E_{\text{PRJ}}(L_{\text{gs}})$].
Note that in Fig.\ 6 the ground-state magic angular momenta obey 
\begin{equation}
L_{\text{gs}} = N(N-1)/2+k(N-1), 
\label{gsam}
\end{equation}
with $k=0$, 1, 2, ... Such sequences, having as a period the number of 
eletrons on the crystalline polygonal ring [5 and 6 for the (1,5) and (1,6) 
RWM's corresponding to $N=6$ and $N=7$], reflect directly the collective 
rotation and incompressibilty of the RWM (see Sect. III.C below).

The stabilization energy, $\Delta E^{\text{gain}}_{\text{gs}}=
E^{\text{gs}}_{\text{PRJ}} - E_{\text{UHF}}$,
of the {\it rotating\/} WM relative to the {\it static\/} 
one [namely the fact that $E^{\text{gs}}_{\text{PRJ}} < E_{\text{UHF}}$,
see Fig.\ 6] is a purely quantum effect. This energy gain,
$\Delta E^{\text{gain}}_{\text{gs}}$, 
demonstrated here for $N=6$ and 7 electrons, is in fact a general property
of states projected out of trial functions with broken symmetry. This is due
to an ``energy gain''  theorem \cite{low} stating that at least one
of the projected states (i.e., the ground state) has an energy lower than
that of the original broken-symmetry trial function (e.g., the UHF determinant
considered above).

\subsection{Excitation gap}

The oscillations of the stabilization energy [Fig.\ (6)] reflect the
oscillatory behavior of the energy of the projected RWM states, as well as
of the exact ones, since the mean-field energy $E_{\text{UHF}}$ varies smoothly
with $B$ (see Ref.\ \onlinecite{yl2}). Underlying the oscillatory behavior
of the ground-state energies is a fundamental property of the spectrum of the
system, namely, the appearance of special gaps due to the enhanced stability
of the cusp states. Indeed, for a given magnetic field, both the ground
state (specified by $L_{\text{gs}}$), as well as the first excited state
(specified by $L_1$), are magic [with \cite{note43} $L_1=
L_{\text{gs}} \pm (N-1)$ for $N=6 - 8$]. As an example of this behavior,
we display in Fig.\ 7 the low part of the EXD excitation spectrum for
a QD with $N=7$, $B=18.8$ T, $\hbar \omega_0 =3$ meV, and $\kappa=12.9$. The 
states $L=99$, $L=105$ (the ground state), and $L=111$ are demonstrated indeed 
to be cusp states of enhanced stability (all three states are well separated 
from the rest of the excited states).

\begin{figure}[t]
\centering\includegraphics[width=6.5 cm]{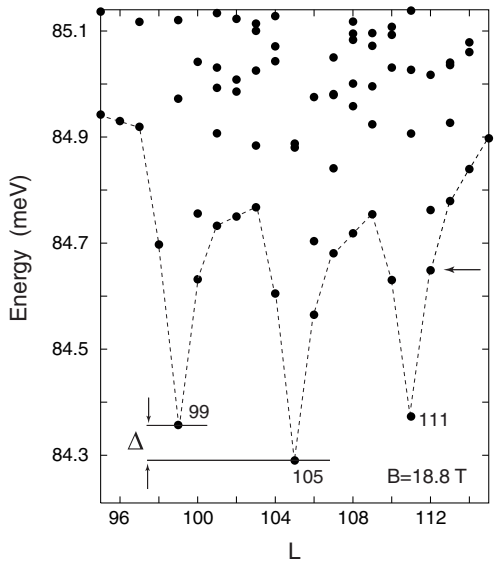}
\caption{
Low-energy part of the spectrum of the parabolic QD whose parameters
are the same as those in Fig.\ 6, calculated as a function of the
angular momentum $L$ through exact diagonalization for $N=7$ electrons 
with a magnetic field $B=18.8$ T. We show here the spectrum in the interval
$95 \leq L \leq 115$ (in the neighborhood of $\nu=1/5$). 
The magic angular momentum values corresponding to cusp states are marked
(99, 105, and 111), and they are seen to be separated from the rest of the 
spectrum. For the given value of $B$, the global energy minimum (ground
state) occurs for $L_{\text{gs}}=105$, and the gap $\Delta$ to the
first excited state ($L=99$) is indicated. The lowest energies for the
different $L$'s in the plotted range are connected by a dashed line, as a 
guide to the eye. The zero of energy corresponds to $7 \hbar \Omega$, where
$\Omega=(\omega_0^2 + \omega_c^2/4)^{1/2}$ and $\omega_c =eB/(m^*c)$.
The horizontal arrow denotes the energy of the Laughlin quasihole at 
$L=112$. It is seen that the Laughlin quasihole is not the lowest
excited state.
}
\end{figure}

\begin{figure}[t]
\centering\includegraphics[width=6. cm]{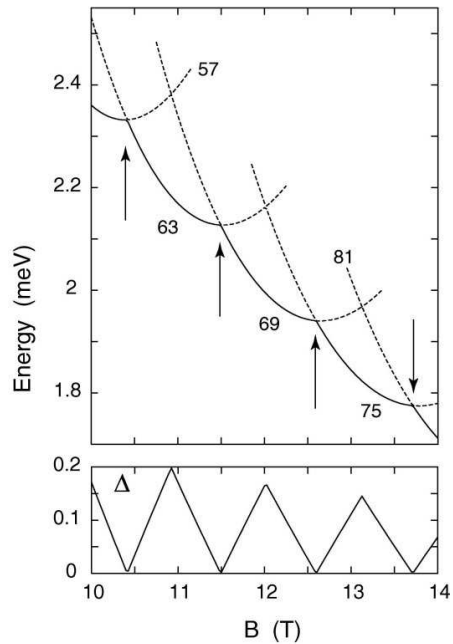}
\caption{Top:
RWM projected energies [see Eq.\ (\ref{eproj})] calculated as a function
of the magnetic field $B$ for $N=7$ electrons in a parabolic QD with
the same parameters as those used in Fig.\ 6 and Fig.\ 7. Each of the
parabola-like curves (made partly of a solid and partly of a dashed line)
corresponds to the marked value of the angular momentum -- i.e., 
for the range of magnetic fields shown here, $L_{\text{gs}}=57$ (7/19),
63 (1/3), 69 (7/23), 75 (7/25), and 81 (7/27), with the corresponding
value of the fractional filling $\nu=N(N-1)/(2L_{\text{gs}})$ given
in parentheses. The solid lines denote the ground-state energies, and the
dashed lines give the values of the first-excited-state energies.
Note that the gap between the ground and the first excited state,
$\Delta = E^1-E^{\text{gs}}$, oscillates as a function of $B$.
The arrows denote the values of $B$ for which the gap vanishes,
occuring between the fractional fillings. The zero of energy corresponds 
to $7 \hbar \Omega + E^{\text{st}}_{\text{cl}}$, where
$\Omega=(\omega_0^2 + \omega_c^2/4)^{1/2}$ [with $\omega_c =eB/(m^*c)$]
and $E^{\text{st}}_{\text{cl}}$ is the classical energy of the 
static Wigner molecule (see Ref.\ \onlinecite{yl2}).
Bottom: The gap $\Delta$ plotted versus $B$.
}
\end{figure}

For the given magnetic field $B=18.8$ T, the first excited state corresponds
to $L=99$. However, as $B$ increases, we found that the state with $L=111$
diminishes in energy relative to that with $L=99$, becoming itself the first 
excited state, and eventually (as $B$ is increased further) replacing $L=105$ 
as the ground state. This sequence
of changes occurs for all ground states (with magic angular momenta) accessed
through variation of the magnetic field. This results in the behavior of
the excitation gap, $\Delta = E^1-E^{\text{gs}}$,
shown in Fig.\ 8 [the calculations here were performed with
the RWM projected energies of Eq.\ (\ref{eproj})]. The gap in Fig.\ 8
separates states with similar internal structure, i.e., they exhibit the
same polygonal configuration as revealed through the CPD analysis for $N=7$
electrons (see Figs.\ 2 $-$ 4). The internal structures of 
higher excited states differ from that of the ground state, with the
disparity increasing with the excitation energy.

The incompressibility of the cusp states (which, as discussed above, 
correspond to magic angular momenta) is connected directly to the appearance
of the gap (as discussed in the context of the FQHE in Ref.\ 
\onlinecite{lau2}). The discussion presented here about the nature of the 
excitation spectrum (and in particular the existence of a gap $\Delta$) allows
us to comment on the influence of impurities and disorder on the properties
of the quantum dot. Naturally, we focus on the regime of small or moderate
disorder, since a high degree of disorder (or strong impurities) will destroy
both the gaps in the spectrum, as well as the coherent (collective) nature of
the rotating Wigner molecule. The effect of disorder (or impurities) depends
on the size of the gap. For sufficiently weak disorder, both the
excitation gap $\Delta$ and the coherence 
of the magic angular momentum states maintain, namely, the states 
separated by the gap experience only {\it local\/}
disorder-induced perturbations (i.e., they broaden) and they remain
conducting (see Sect. IV below). Obviously, for cases of a vanishing gap
(i.e., between the fractional fillings, see arrows in Fig.\ 8), even 
weak disorder can induce a {\it global\/} change in the character of the
(perturbed) wave functions by strongly mixing the degenerate $L_{\text{gs}}$
and $L_1$ cusp states (and often additional nearby cusp states depending
on the magnitude of $B$), and this can lead to a state with a broken-symmetry
electron density having characteristics of a pinned Wigner 
crystallite. \cite{ylunp}

\section{Discussion: Implications for the thermodynamic limit}

While our focus here is on the behavior of trial and exact wave functions
in (finite) QD's in high magnetic fields, it is natural to inquire about
possible implications of our findings to FQHE systems in the thermodynamic
limit. 

We recall that appropriate trial wave functions for clean FQHE systems
possess a good angular momentum $L \geq L_0$, a property shared by
both the CF/JL and RWM functions. \cite{yl4b,lau2,lau3,jai2} 
We also recall the previous finding \cite{note75} 
that for large fractional fillings $\nu > \nu_0$, the liquid-like (and 
circularly uniform) CF/JL function is in the thermodynamic limit 
energetically favored compared 
\cite{lau2,lau3,jai4,lam} to the broken-symmetry static Wigner crystal (which 
has no good angular momentum); for $\nu < \nu_0$, the static Wigner crystal 
becomes lower in energy. This finding was enabled by the form of the JL 
functions, which facilitated computations of total energies as a function of 
size for sufficiently large $N$ (e.g., $N=1000$).

A main finding of this paper is that the {\it exact-numerical-diagonalization 
wave functions of small systems ($N \leq 10$) are crystalline in character 
for both low and high fractional fillings.\/} 
This finding contradicts earlier suggestions \cite{lau2,lau3,jai1,jai3} that,
for high $\nu$'s, small systems are accurately described by the liquid-like
JL wave functions and their descendants, e.g., the composite-fermion ones. 
Of course, for the same high $\nu$'s,
our small-size results cannot exclude the possibility that the CPD's of the 
exact solution may exhibit with increasing $N$ a transition from crystalline 
to liquid character, in agreement with the JL function. 
However, at the moment the existence of such a transition remains an open
theoretical subject.

For the {\it low fractions\/}, the RWM theory raises still 
another line of inquiry.
Due to the specific form of the RWM wave functions, computational limitations
\cite{note} prevent us at present from making extrapolations of total energies
at a given $\nu$ as $N \rightarrow \infty$. Nevertheless, from the
general theory of projection operators, one can conclude that the RWM
energies exhibit a different trend compared to the JL ones, whose energies
were found \cite{lau2,lau3,jai4,lam} to be higher  
than the static Wigner crystal. Indeed the 
rotating-Wigner-molecule wave functions remain lower in energy than the 
corresponding {\it static\/} crystalline state for {\it all values\/} of $N$ 
and $\nu$, even in the thermodynamic limit. This is due to the fact that the 
aforementioned energy-gain theorem \cite{low} (see sect. III) applies for 
any number of electrons $N$ and for all values of the magnetic field $B$.
Naturally, the RWM wave functions will be physically relevant compared to 
those of the broken-symmetry crystal at the thermodynamic limit if 
the energy gain does not vanish when $N \rightarrow \infty$; otherwise, one 
needs to consider the posssibility that the static crystal is the relevant 
physical picture.

The discussion in the above paragraph may be recapitulated by the following 
question: which state is the relevant one in the thermodynamic limit
$(N \rightarrow \infty)$ -- the broken-symmetry one (i.e., the static Wigner
crystal) or the symmetry restored (i.e., rotating crystal) state? This
question, in the context of bulk broken-symmetry systems, has been addressed
in the early work of Anderson, \cite{pwa} who concluded that 
the broken-symmetry state (here the UHF static crystalline solution)
can be safely taken as the effective ground state. In arriving at this
conclusion Anderson invoked the concept of (generalized) rigidity.
As a concrete example, one would expect a crystal to behave like a 
{\it macroscopic\/} body, whose Hamiltonian is that of a {\it heavy rigid 
rotor\/} with a low-energy excitation spectrum $L^2/2{\cal J}$, the moment of
inertia ${\cal J}$ being of order $N$ (macroscopically large when 
$N \rightarrow \infty$). The low-energy excitation spectrum of this heavy 
rigid rotor above the ground-state ($L=0$) is essentially gapless (i.e., 
continuous). Thus although the formal ground state posseses continuous
rotational symmetry (i.e., $L=0$), ``there is a manifold of other states, 
degenerate in the $N \rightarrow \infty$ limit, which can be recombined to 
give a very stable wave packet with essentially the nature'' \cite{note34}
of the broken-symmetry state (i.e., the static Wigner crystal in our case).
As a consequence of the ``macroscopic heaviness'' as $N \rightarrow \infty$, 
one has: (I) The energy gain due to symmetry restoration (i.e., the 
stabilization energy $\Delta E^{\text{gain}}_{\text{gs}}$) vanishes as 
$N \rightarrow \infty$, and (II) The relaxation of the system from the wave 
packet state (i.e., the static Wigner crystal) to the symmetrized one (i.e., 
the rotating crystal) becomes exceedingly long. This picture underlies the 
aforementioned conclusion that in the thermodynamic limit the broken-symmetry 
state may be used as the effective ground state.

Consequently, in the following we will focus on issues 
pertaining to the ``rigidity'' of the rotating Wigner molecule
in high magnetic fields. In particular, using our projection method and exact 
diagonalization, we demonstrated explicitly in previous publications 
\cite{yl2,yl5} that the rigid-rotor picture applies to an $N$-electron QD only
when $B=0$. In contrast, in the presence of a high 
magnetic field, we found \cite{yl2} that the electrons in the QD do not 
exhibit global rigidity and therefore cannot be modeled as a macroscopic
rotating crystal. Instead, a more appropriate model is that
of a {\it highly non-rigid\/} rotor whose moment of inertia
depends strongly on the value of the angular momentum $L$. 

The non-rigid rotor at high $B$ has several unique properties: 
(I) The ground state has angular 
momentum $L_{\text{gs}} > 0$; (II) While the rotating electron molecule 
does not exhibit {\it global\/} rigidity, it possesses {\it azimuthal\/} 
rigidity (i.e., all electrons on a given ring rotate coherently), with the 
rings, however, rotating independently of each other.
Furthermore, the radii of the rings vary for different values of $L$,
unlike the case of a rigid rotor (see for example the locations of ${\bf r}_0$
in Fig.\ 1 and Fig.\ 3 for different $L$ values);
(III) The excitation spectra do not vary as $L^2$; instead they consist of 
terms that vary as $aL+b/\sqrt{L}$ [for the precise values of the constants 
$a$ and $b$ in the case of $(0,N)$ or $(1,N-1)$ electron molecules, 
see Ref.\ \onlinecite{yl2}]; (IV) The angular momentum values are given by the
magic values [see Ref.\ \onlinecite{yl4b}] $L=L_0+\sum_{q=1}^r k_q n_q$, where 
$(n_1,n_2,...,n_r)$ is the polygonal ring arrangement of the static Wigner 
molecule (with $n_q$ the number of electrons on the $q$th
ring) and $k_1 < k_2 <...< k_r$ are nonnegative integers. 
These magic $L$'s are 
associated with the cusp states which exhibit a relative energy gain with
respect to neighboring excitations. Thus the low-energy excitation spectrum
of the non-rigid rotor is not dense and exhibits gaps due to the occurrence of
the magic (cusp) states (see section III.C). Furthermore, these gaps are 
reflected in the oscillatory behavior of $\Delta E^{\text{gain}}_{\text{gs}}$ 
(see, e.g., Fig.\ 6) as a function of $B$ (or $\nu$).

As $N$ increases, more polygonal rings are successively added, and since 
the polygonal rings rotate independently of each other
(see, e.g., the case of $N=9$ in Fig.\ 5), we expect that the
non-rigid-rotor picture remains valid even as $N \rightarrow \infty$. As a 
result, it is plausible to conjecture the following properties at high $B$ in
the thermodynamic  limit: 
(I) the oscillatory character of $\Delta E^{\text{gain}}_{\text{gs}}$ will 
maintain, yielding nonvanishing  stabilization energies at the fractional
fillings $\nu$ [see Eq.\ (\ref{nu})], and (II) the low-energy excitation 
spectra of the system will still exhibit gaps in the neighborhood of the magic
angular momenta (see Fig.\ 8). 
Of course, these conjectures need to be further supported 
through numerical calculations for large $N$. Nevertheless, 
the above discussion indicates that the question of which state is physically 
relevant for low fractions in the thermodynamic limit at high $B$ -- i.e., the 
broken-symmetry static crystal or the symmetrized rotating crystal -- 
remains open, and cannot be answered solely following the path
of Ref.\ \onlinecite{pwa}.

The rotating Wigner crystal has properties characteristic of FQHE 
states, i.e., it is incompressible (connected to the presence of an 
excitation gap) and carries a current \cite{sond} 
(while the broken-symmetry static crystal is insulating). 
Thus, we may conjecture that a transition at lower fractional fillings from a 
conducting state with good circular symmetry to an insulating Wigner crystal 
cannot occur {\it spontaneously\/} for clean systems. Therefore, it should be 
possible to observe FQHE-type behavior at low fractional fillings in a clean 
system -- a prediction that could explain the observations of Ref.\ 
\onlinecite{pan}, where FQHE behavior has been observed for low 
fractional fillings typically associated with the formation of a
static Wigner crystal. In practice, however, impurities 
and defects may influence the properties of the rotating crystal (and its 
excitations), depending on the magnitude of the excitation gap discussed in 
the previous section. Thus one of the main challenges for FQHE observation at 
such low fillings relates to fabrication of high mobility (nearly 
impurity-free) samples. \cite{wes,note2} 

\section{Summary}

In summary, we have carried out the first systematic investigations
(for $6 \leq N \leq 9$) of structural properties of 
cusp states in parabolic quantum dots at high 
magnetic fields. Our anisotropic conditional probability distributions from 
exact diagonalization show that these states are crystalline in character for 
both low and high fractional fillings, unlike the liquid-like Jastrow-Laughlin 
\cite{lau2,lau3} wave functions, but in remarkable agreement with the recently 
proposed rotating-Wigner-molecule \cite{yl1,yl4b} ones. The cusp states of 
$N$-electron parabolic QD's are precursors to the extended fractional quantum 
Hall states (and not to the static Wigner crystal) due to 
stabilization of the {\it rotating\/} Wigner molecule (having a good
angular momentum) relative to the {\it static\/} one (that exhibits broken 
symmetry). The rotating Wigner molecule in high $B$ does not exhibit 
global rigidity; instead, it possesses {\it azimuthal\/}
rigidity (i.e., all electrons on a given ring rotate coherently), with the
rings, however, rotating independently of each other.

Furthermore, we demonstrated pertinent features of the spectrum of
quantum dots in high $B$, showing that
both the ground state and the first excited state correspond to magic
angular momenta (cusp states). For a given $B$, this leads to the
appearance of a special gap that is not a mere consequence of the finite
size of the system (and thus it is expected to maintain in the thermodynamic
limit, underlying the incompressibility of the electron system).
Finally, we discussed in detail issues pertaining 
to the implications of the rotating-Wigner-molecule theory for FQHE systems
at the thermodynamic limit (see sect.\ IV).

We thank M. Pustilnik for comments on the manuscript.
This research is supported by the U.S. D.O.E. (Grant No. FG05-86ER45234).


\begin{thebibliography}{99}
\bibitem{note11}
In previous publications the RWM wave functions were also
referred to as rotating-electron-molecule (REM) wave functions.
\bibitem{note3}
The results in Figs.\ 1-5 were obtained without consideration of 
a confinement, and they are not influenced (high $B$) by inclusion 
of a parabolic confinement [Refs.\ \onlinecite{jai1,yang,yl2}]. 
For the influence of the confinement on the energies, 
see Refs.\ \onlinecite{jai1,yang,yl2}.
\bibitem{note43}
This expression for $L_1$ corresponds to the $(1,N-1)$ arrangement
of the $N$-electron Wigner molecule. The $(0,N)$ arrangement is found to be the
global ground state for a limited range of values of the fractional filling 
factor $\nu$ near $\nu=1$.
\bibitem{note75}
Several different values for $\nu_0$ have been proposed in the
literature. Ref.\ \onlinecite{lau2} determined a value $\nu_0=1/11$.
Ref.\ \onlinecite{lam} gave a value of $\nu_0=1/7$. Recently,
Ref.\ \onlinecite{jai4} argued in favor of the value $\nu_0=1/11$.
\bibitem{note}
We remark that, at present, computational difficulties prohibit
a straightforward extrapolation of energy to the thermodynamic 
limit as a function of $N$ and $\nu$. This is the case especially for the disk
geometry used here, which is advantageous for studies of crystalline
structures, compared to the computationally easier spherical geometry.
\bibitem{note2}
We remark, however, that the stabilization energy and the gap $\Delta$ 
diminish as the magnetic field increases (see Fig.\ 6 and Fig.\ 8), and as a 
result the impurities become more efficient in influencing the WM for the 
lower fractional fillings (i.e., higher angular momenta).
\end{thebibliography}
\end{document}